\documentclass[
    ,final           ]
  {aipproc}

\layoutstyle{8x11single}

\begin{document}

\title{Search for Extra Dimensions at CDF}


\author{S-M Wynne}{
  address={University of Liverpool, UK}
}

\begin{abstract}
This poster, presented at the 2006 Duke Hadron Collider Symposium, presents the results from searches for large extra dimensions, as proposed by Arkani-Hamed, Dimopoulos and Dvali (ADD), and Randall-Sundrum (RS) model warped extra dimensions, at CDF. 
 \end{abstract}

\maketitle

%%%%%%%%%%%%%%%%%%%%%%%%%%%%%%%%%%%%%%%%%%%%
%% MAINMATTER
%%%%%%%%%%%%%%%%%%%%%%%%%%%%%%%%%%%%%%%%%%%%

\section{Introduction}
Through lowering the unification of the forces from the GUT scale to the weak scale, the fine tuning cancellation required due to heavy gauge bosons coupling to the Higgs may be removed through lowering the Planck scale ($M_{Pl}$), where gravity becomes strong, thereby increasing gravity's true strength. To maintain the perceived strength that we observe, gravity is assumed to leave the (3+1) space-time dimensions of the Standard Model (SM) brane and propagate into the extra dimensions of the bulk, resulting in a reduced Planck scale ($M_{D}$), the fundamental scale in space-time, with only the weaker projection of gravity being observable on the SM brane. \\
Large extra dimensions (LEDs), proposed by Arkani-Hamed, Dimopoulos and Dvali (ADD)~\cite{ADD}, consider the scenario in which the SM particles and gauge interactions are confined to the 3-brane which is embedded in a (3 + $n$) dimensional bulk, where $n$ are assumed compact to preserve Newtonian gravity on a large scale. The compactification radius $R$ is given by $R^{n} = M_{Pl}^{2} / (8\pi M_{D}^{n+2})$. Gravitons are able to propagate in the bulk, appearing as a tower of Kaluza-Klein $($KK$)$ excited modes on the SM brane, and can be detected via virtual graviton exchange, leading to a modification in dilepton or diphoton cross section, or direct graviton emission detectable as single jets or photons with missing transverse energy.\\
Randall and Sundrum~\cite{RS} proposed an alternative solution with a single extra dimension with a non-Euclidian (AdS) warped metric with curvature $k$ extending between the Planck brane, where gravity is localized, and the TeV brane where the SM fields are localized. This metric leads to the graviton wave function being exponentially suppressed away from the brane along the extra dimension with a scale of order $\Lambda_{\pi} = \bar{M}_{Pl}exp(-k\pi R)$, where $\bar{M}_{Pl}   \equiv M_{Pl} / \sqrt{8\pi}$ is the reduced Planck mass. Setting $\Lambda_{\pi} \sim 1$~TeV then produces the desired fundamental Planck scale. Gravitons are then free to propagate in the bulk and are detectable as KK modes on the TeV brane. The lowest order of these KK modes $M_{0}$ is massless, thus a limit is put on $M_{1}$, the mass of the first resonance.\\

\section{Search for Large Extra Dimensions}

\subsection{Virtual graviton exchange}
\begin{figure}[htbp]
\includegraphics[height=.2\textheight]{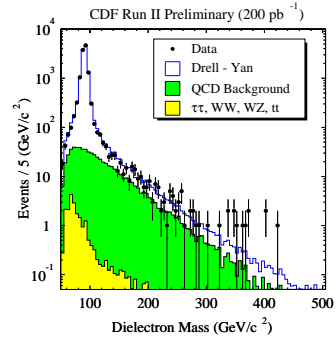}
\caption{Dielectron mass distribution showing data and SM background}
\label{fig:e_spectrum}
\end{figure}

\begin{figure}[htbp]
\includegraphics[height=0.1\textheight]{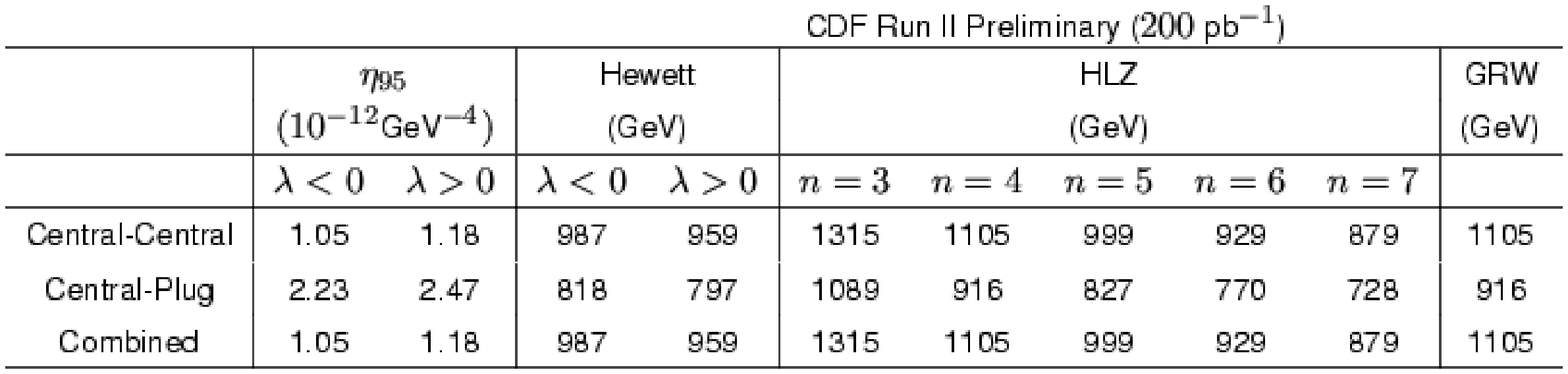}
\caption{Effective Planck scale limits for virtual graviton exchange in the dielectron channel.}
\label{fig:limits_ee}
\end{figure}

The cross section is modified by the term $d^{2}\sigma / dMdcos\theta^{*} = f_{SM} + f_{int}\eta_{G} + f_{KK}\eta_{G}^{2}$, where $f_{SM}$ is the SM term, $f_{int}$ the interference and $f_{KK}$ the graviton terms, $M$ is the invariant dilepton or diphoton mass and $\theta^{*}$ is the scattering angle. Several formalisms are used for the effective Lagrangian; GRW~\cite{GRW}, HLZ~\cite{HLZ} and Hewett~\cite{Hewett} and accounted for through the dimensionless parameter $\lambda$, where $\eta_{G} = \lambda/M^{4}_{s}$ and $M_{s}$ represents the cutoff required to keep the sum of KK states finite and is of the order $M_{D}$.\\
This search  is currently performed in the dielectron channel only~\cite{die_led_virt} using 200~$\rm{pb^{-1}}$ of data. Two isolated electrons were required, with transverse energy $\rm{E_{T}}$ > 25 GeV and located in the central or forward regions of the calorimeter, with at least one electron in the central region. The invariant mass spectrum for both regions is shown in Figure~\ref{fig:e_spectrum} and is in agreement with the expected background.  The effective Planck scale limits for the three formalisms and lambda conventions ($\rm{\lambda \pm = 1}$) are shown in Table~\ref{fig:limits_ee}.

\subsection{Direct graviton emission}
\begin{figure}[htbp]
 \includegraphics[height=.2\textheight]{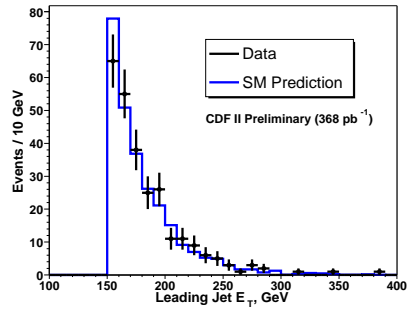}
 \caption{Comparison of leading jet $\rm{E_{T}}$ distribution from predicted events from SM sources and events seen in data. }
 \label{et1}
 \end{figure}

\begin{figure}[htbp]
 \includegraphics[height=.2\textheight]{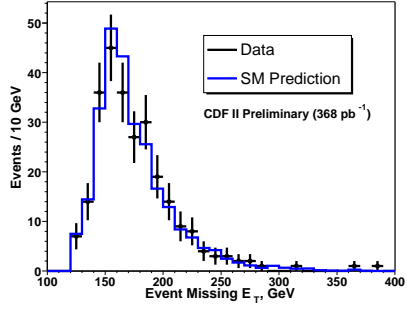}
 \caption{Comparison of missing $\rm{E_{T}}$ distribution from predicted events from SM sources and events seen in data. }
\label{met}
\end{figure}

LEDs may be detected through gravitons which are produced recoiling against a quark or a gluon jet, resulting in a jet plus missing $\rm{E_{T}}$ signature. This analysis~\cite{direct} uses 368~$\rm{pb^{-1}}$ of data, collected with a high $\rm{E_{T}}$ single jet trigger with the highest $\rm{E_{T}}$ jet in each event required to be in the central region of the detector. Each reconstructed jet with $\rm{E_{T}}$ > 20 GeV is required to have matching reconstructed tracks whose sum $\rm{p_{T}}$ corresponds to at least 10\% of the jet $\rm{E_{T}}$ and missing $\rm{E_{T}}$ > 80 GeV is asked for.  A threshold of 150 GeV is then placed on the leading jet to select the final events from the monojet candidate sample. Figures~\ref{et1} and~\ref{met} show the comparison of predicted events from SM MC and events seen in data, which are in agreement.  Based on the maximum possible number of observed signal events, limits are placed on $\rm{M_{D}}$, shown in Table~\ref{tab:metlim}, for K factors (the expected ratio of cross sections at next to leading order and leading order) of 1.0 and 1.3. These can be related to limits on the radius of the extra dimensions via the expression $R^{n} = 1/8\pi (M_{Pl}/M_{D}) 1/M_{D} ^ {n}$. Similarly, direct graviton emission may lead to a photon and missing $\rm{E_{T}}$ signature. For this search one photon was required with $\rm{E_{T}}$ > 55 GeV and to be located in the central region of the detector. Missing $\rm{E_{T}}$ of > 45 GeV was also required, with no jets with $\rm{E_{T}}$ > 15 GeV and no tracks with $\rm{p_{T}}$ > 5 GeV.  72~$\rm{pb^{-1}}$ of data were used to set limits on $M_{D}$, also shown in Table~\ref{tab:metlim}.

\begin{table}
\begin{tabular}{cccc}
\hline
  \tablehead{1}{c}{b}{n\\}
 & \tablehead{1}{c}{b}{$\rm\mathbf{{M_{D} (K = 1.0)}}$\\ jet + MET}
  & \tablehead{1}{c}{b}{$\rm\mathbf{{M_{D} (K = 1.3)}}$\\ jet + MET} 
  & \tablehead{1}{c}{b}{$\rm\mathbf{{M_{D} (TeV)}}$\\ photon + MET} \\
  \hline
2 & 1.09 & 1.16 & - \\
3 & 0.93 & 0.98 & - \\
4 & 0.87 & 0.90  & > 0.55\\
5 & 0.82 & 0.85  & - \\
6 & 0.80 & 0.83 & > 0.58 \\
8 &  -       &  -  & > 0.60 \\
\hline
\end{tabular}
\caption{95 \% confidence level lower limit on $\rm{M_{D}}$ for n = 2-6 with K factors of 1.0 and 1.3 for jets plus missing $\rm{E_{T}}$ and limits on $\rm{M_{D}}$ for n = 4-8 for photons plus missing $\rm{E_{T}}$ .}
\label{tab:metlim}
\end{table}

\section{Search for Warped Extra Dimensions}
Virtual graviton exchange is considered for RS model searches, where the excited KK modes can decay into fermion-antifermion or diboson pairs and be detected as a resonance in the invariant mass spectrum, or cause a modification in angular distribution. The analyses presented here consider the dielectron~\cite{die_led_virt} ($\rm{200 pb^{-1}}$), diphoton~\cite{rs_pho} ($\rm{345 pb^{-1}}$) and dimuon~\cite{rs_mu} ($\rm{200 pb^{-1}}$) final states. Two isolated candidates are required. Electrons require an EM cluster with $\rm{E_{T}}$ > 25 GeV and photons with $\rm{E_{T}}$ > 15 GeV. All muon candidates are required to have a COT track with $\rm{p_{T}}$ > 20 $\rm{GeV/c^{2}}$. The data are in agreement with the expected background and 95\% confidence level upper limits are set on the cross section times branching ratio ($\sigma \cdot BR(\gamma \gamma/ ll)$), allowing lower  limits to be set on the graviton mass as a function of $k/M_{Pl}$, as shown in Table~\ref{tab:mass}.

\begin{table} 
\begin{tabular}{lcccc}
\hline
  \tablehead{1}{c}{b}{Channel\\}
 & \tablehead{1}{c}{b}{Luminosity\\$\rm\mathbf{{pb^{-1}}}$}
  & \tablehead{1}{c}{b}{$\rm\mathbf{{M_{G} (GeV/c^{2})}}$\\$\rm\mathbf{{k/M_{Pl} = 0.01}}$}
   & \tablehead{1}{c}{b}{$\rm\mathbf{{M_{G} (GeV/c^{2})}}$\\$\rm\mathbf{{k/M_{Pl} = 0.1}}$}   \\
\hline
ee & 200 & 200 & 640 \\
$\mu\mu$ & 200 & 170 & 610 \\
ee + $\mu\mu$ & 200 & 200 & 700  \\
$\gamma \gamma$ & 345 & 220 & 690  \\
\hline
\end{tabular}
\caption{Lower limit on RS graviton mass for dilepton and diphoton channels.}
\label{tab:mass}
\end{table}

 %%%%%%%%%%%%%%%%%%%%%%%%%%%%%%%%%%%%%%%%%%%%%%%%
%% BACKMATTER
%%%%%%%%%%%%%%%%%%%%%%%%%%%%%%%%%%%%%%%%%%%%%%%%
\clearpage
\bibliographystyle{aipproc}

\end{document}